\documentclass[aps,prd,preprint,nofootinbib]{revtex4}  % for review and submission
\usepackage{latexsym,amsmath,amssymb,amstext}
\usepackage{slashed}
\usepackage{epsfig}
\usepackage{float,xcolor}
\usepackage{graphicx}
\usepackage{dcolumn}

\newcommand{\be}{\begin{equation}}
\newcommand{\ee}{\end{equation}}
\newcommand{\bea}{\begin{eqnarray}}
\newcommand{\eea}{\end{eqnarray}}
\newcommand\bef{\begin{figure*}}
\newcommand\eef[1]{\label{fg:#1}\end{figure*}}
\newcommand\beq{\begin{equation}}
\newcommand\eeq[1]{\label{#1}\end{equation}}
\newcommand\beqa{\begin{eqnarray}}
\newcommand\eeqa[1]{\label{#1}\end{eqnarray}}
\newcommand\bet{\begin{table}}
\newcommand\eet[1]{\label{tb:#1}\end{table}}

\newcommand\fgn[1]{Figure \ref{fg:#1}}
\newcommand\eqn[1]{Eq.\ (\ref{#1})}

\usepackage[paperwidth=210mm,paperheight=297mm,centering,hmargin=2cm,vmargin=2.5cm]{geometry}

\begin{document}

\widetext

\title{Numerical determination of monopole scaling dimension in
parity-invariant three-dimensional non-compact QED}

\author{Nikhil\ \surname{Karthik}}
\email{nkarthik@bnl.gov}
\affiliation{Physics Department, Brookhaven National Laboratory, Upton, New York 11973-5000, USA}
\author{Rajamani\ \surname{Narayanan}}
\email{rajamani.narayanan@fiu.edu}
\affiliation{Department of Physics, Florida International University, Miami, FL 33199}

\begin{abstract}
We present a direct Monte-Carlo determination of the scaling dimension
of a topological defect operator in the infrared fixed point of a
three-dimensional interacting quantum field theory. For this, we
compute the free energy to introduce the background gauge field of
the $Q=1$ monopole-antimonopole pair in three-dimensional non-compact
QED with $N=2,4$ and $12$ flavors of massless two-component fermions,
and study its asymptotic logarithmic dependence on the
monopole-antimonopole separation. We estimate the scaling dimension
in the $N=12$ case to be consistent with the large-$N$ (free fermion)
value. We find the deviations from this large-$N$ value for $N=2$
and $4$ are positive but small, implying that the higher order
corrections in the large-$N$ expansion become mildly important for
$N=2,4$.
\end{abstract}

\maketitle

\section{Introduction}

Conformal field theories in three-dimensions, and renormalization
group flows from one fixed point to another induced by the introduction
of relevant operators at fixed points have been investigated over
the last few years. This involves the computation of scaling
dimensions, $\Delta$, of operators at the different fixed points.
The operators ${\cal O}$ at a fixed point could be the usual
composites of the field variables, and hence trivially local and
amenable to the standard Monte Carlo computations of two-point
functions of the local operator
\beq
\left\langle {\cal O}(x) {\cal O}(0)\right\rangle \sim \frac{1}{|x|^{2\Delta}}.
\eeq{oscaledim}
The operators could also be topological disorder
operators~\cite{Borokhov:2002ib} which act as sources to topological
conserved currents in the theory.  Since such operators cannot be
written as simple composites of the field variables, studying their
scaling dimensions is a challenge, especially on the numerical side.
In the case of theories with U$(1)$ global or gauged symmetry, the
topological defects are the monopoles, $M_Q$, which create $Q$ units
of flux surrounding it~\cite{Borokhov:2002ib}, and hence serve as
the sources of the otherwise trivially conserved U$_{\rm top}(1)$
current, $j_\mu^{\rm top}=\epsilon_{\mu\nu\rho}F^{\nu\rho}/(4\pi)$.
Three-dimensional QED, whose gauge group is U$(1)$ as opposed to
${\mathbf R}$, is one such theory where monopole defects can occur.
Depending on whether monopoles are energetically allowed or disallowed
in the continuum limit, the three-dimensional QED is classified as
compact or non-compact respectively. The presence of two distinct
theories, differing simply by the presence or absence of monopoles,
offers theoretical and computational possibilities in understanding
the emergence of mass-gap.

Pure-gauge compact QED in three-dimensions is a rare example in
which the emergence of mass-gap could be understood through the
dual superconductor mechanism where in the electric charges experience
a linear confining potential due to the presence a plasma of magnetic
monopoles~\cite{Polyakov:1975rs,Polyakov:1976fu}. Coupling the
compact QED$_3$ (referred to as c-QED$_3$) to many flavors, $N$, of
massless two-component fermions (assumed to be even to preserve
parity) gives a possibility to counter the emergent
mass-gap~\cite{Pufu:2013vpa} --- above certain critical flavor $N^{\rm
C}_c$, the theory is expected to be conformal in the infra-red,
whereas develops a mass-gap below $N^{\rm C}_c$. 
In a first exploratory study~\cite{Hands:2006dh} towards finding $N^{\rm C}_c$,
a derivative of free-energy required to introduce a single 
monopole was computed on a lattice with open boundary conditions in the
$N=8$ flavor compact QED$_3$, and no convincing evidence for an infinite 
free-energy signalling monopole-confinement was found in the continuum
limit. Thereby,
this suggested a presence of 
monopole plasma and a consequent mass-gap in compact QED$_3$ for $N\le 8$.
Understanding such
infra-red quantum phases obtained by tuning parameters of the
underlying QFT is an ongoing field of research (c.f.,~\cite{Wang:2017txt}).
Similar studies of the critical number of flavors, $N_c^{\rm NC}$,
in non-compact QED$_3$ (referred to as nc-QED$_3$) is continued to
be investigated through ab initio lattice
simulations~\cite{Hands:2002dv,Hands:2004bh,Raviv:2014xna,Karthik:2015sgq,Karthik:2016ppr}
as well as through other approximation
methods~\cite{DiPietro:2015taa,Giombi:2015haa,Kotikov:2016wrb,Gusynin:2016som,DiPietro:2017kcd,Herbut:2016ide}.
Unlike non-compact QED$_3$, the presence of monopoles in the compact
version even as the continuum limit is approached, is a technical
challenge to numerical studies due to the presence of many small
eigenvalues of the three-dimension Dirac operator~\cite{Armour:2011zx}.
An indirect feasible approach is to check whether the monopole
operator is marginally relevant in the infrared fixed point of  the
$N$ flavor non-compact QED$_3$~\cite{Pufu:2013vpa}.  Crucial to
this inference is that the monopoles in a gauge theory with $N$
massless fermions break ${\rm U}(N)$ global flavor symmetry to ${\rm
U}(N/2)\times {\rm U}(N/2)$
symmetry~\cite{Pufu:2013vpa,Dyer:2013fja,Chester:2015wao}.  Such
an approach further assumes that 1) both compact and non-compact
QED$_3$ flow to the same infrared fixed point for $N>N^{\rm C}_c$;
2) $N_c^{\rm NC} < N^{\rm C}_c$.  At least in the $N\to\infty$
limit, the compact or non-compact action will be sub-dominant
compared to the induced gauge action from the fermion, and hence,
the infrared physics should be the same for both nc- and c-QED$_3$.
The stronger assumption is that this continues to remain so until
$N=N^{\rm C}_c$. The second assumption is based more on numerical
works~\cite{Karthik:2015sgq,Karthik:2016ppr} that strongly indicate
that $N_c^{\rm NC} < 2$. This also means that only the dressed,
gauge-invariant monopole operators become relevant at $N=N^{\rm
C}_c$ and other U$(N)$ symmetry breaking operator, such as the
four-fermi operators, remain irrelevant~\cite{Chester:2016ref}.
Therefore, a computation of scaling dimensions in nc-QED$_3$ and a
subsequent direct confirmation of $N^{\rm C}_c$ in c-QED$_3$ is
well motivated.  
Monopole operators also play similar role to
understand quantum phase transitions in lattice systems with gauged
U$(1)$ symmetry which was recently analyzed computationally in
compact QED$_3$~\cite{Xu:2018wyg,Wang:2019jxs}, and in QED$_3$-Gross-Neveu
model~\cite{Dupuis:2019uhs}.  In~\cite{Song:2018ial}, analytical
progress was made on monopoles in such lattice systems.

A practical method to determine the monopole scaling dimension
$\Delta_Q$ analytically is by coupling the theory with the U$(1)$
symmetry to the classical, scale- and rotationally-invariant Dirac
monopole background ${\cal A}^Q$ and study the response of the
theory.  Analytically, one computes the Casimir energy of the theory
defined on $S^2$ with uniform $2\pi Q$ flux over it, which by
state-operator correspondence is the same as the scaling dimension
$\Delta_Q$~\cite{Murthy:1989ps,Metlitski:2008dw,Pufu:2013eda}.  Such
computations are usually perturbatively done order by order in $1/N$
(c.f.,~\cite{Gracey:2018ame}), and currently it is only up to $O(1/N)$.
Non-perturbative conformal bootstrap has also been applied to QED$_3$
to find the allowed region in the parameter space of scaling
dimensions of $Q=1$ and 2 monopoles~\cite{Chester:2016wrc}.
Complementary to such bootstrap computations, it was
demonstrated~\cite{Karthik:2018rcg} that a direct way to compute
monopole scaling dimensions using lattice computation is to couple
such theories to a background field ${\cal A}^{Q\overline{Q}}(x;\tau)={\cal
A}^Q(x;x_0)-{\cal A}^Q(x;x_0+\hat{t}\tau)$ that gives rise to a
monopole at $x_0$ and an anti-monopole at $x_0+\hat{t}\tau$, which
are separated by a distance $\tau$ and compute the scaling of the
partition function
\beq
Z(A^{Q\overline{Q}}(\tau))\sim \frac{1}{\tau^{2\Delta_Q}},
\eeq{monscale}
as $\tau\to\infty$. It is the aim of this paper to apply this method
and  compute $\Delta_Q$ for $Q=1$ monopole in the infrared fixed
points in $N$ flavor noncompact QED$_3$. In particular, we compute
the finite $N$ corrections to the large-$N$ scaling dimension for
small enough values of $N$ where a nonperturbative computation
becomes inevitable.

\section{c-QED$_3$, nc-QED$_3$ and monopole correlator in nc-QED$_3$}

In this section, we consider different versions of QED$_3$ that one
could construct on the lattice.  We consider $L^3$ Euclidean lattices
whose physical volume is $\ell^3$, with the lattice spacing being
$\ell/L$.  Let $\theta_\mu(x) \in {\mathbf R}$ be the lattice gauge
fields which are related to the physical gauge fields
$\theta_\mu(x)=A_\mu(x) \ell/L$. The notation is such that $x$,$y$
denote integer valued lattice coordinates. The two-component Dirac
fermions in all the cases to be considered, are coupled to compact
gauge-links, $U_\mu(x)=e^{i\theta_\mu(x)}$, through a UV regulated
massless Dirac operator $\slashed{C}(U)$. In this work, $\slashed{C}(U)$
is the 1-HYP smeared Wilson-Sheikholeslami-Wohlert Dirac operator
$\slashed{C}_W$ with the Wilson mass $m_w$ tuned to the massless
point~\cite{Karthik:2015sgq}.  In the parity-invariant QED$_3$ with
even number of flavors,  $N/2$ of two-component fermions are coupled
via $\slashed{C}(U)$ and the other $N/2$ via $\slashed{C}^\dagger(U)$.
The partition function for QED$_3$ can be written in general as
\beq
 Z=\left(\prod_{x,\mu}\int_{-\infty}^\infty d\theta_\mu(x)\right) {\det}^{N/2}\left[\slashed{C}^\dagger(U) \slashed{C}(U)\right] \times
 {\cal W}_{\rm g},
\eeq{genact}
where ${\cal W}_g$ is the Boltzmann weight from the pure gauge part.
Since the fermionic determinant is invariant under $\theta_\mu(x)
\to \theta_\mu(x) + 2\pi n_\mu(x)$ for integer values $n_\mu(x)$,
this part of the action respects the compactness of the U(1) gauge
group.  Independent of the choice of ${\cal W}_g$, we can always
restrict the above integral over all $\theta_\mu(x)$ to be from
$-\pi$ to $\pi$ by simply summing ${\cal W}_g$ over all possible
$n_\mu(x)$ for different $x$ and $\mu$.  In this way, the underlying
gauge group is always U(1) owing to the usage of the compact links
$U_\mu(x)$ in the Dirac operator, and hence magnetic monopoles are
well defined in these theories.  Depending on the form of ${\cal
W}_{\rm g}$, one can study QED$_3$ with or without monopoles as we
elaborate below, and also discussed in~\cite{Sulejmanpasic:2019ytl}.

All gauge actions will be functions of the fluxes on plaquettes
where the flux on the plaquette in the $(\mu,\nu)$ plane with one
corner at $x$ is
\be
F_{\mu\nu}(x)=\nabla_\mu\theta_\nu(x)-\nabla_\nu\theta_\mu(x);\qquad \nabla_\mu f(x) = f(x+\hat\mu) - f(x).
\ee
The Boltzmann weight for the non-compact lattice gauge action,
\beq
{\cal W}_{\rm g}\equiv e^{-\sum_x S_{\rm nc}(x)};\qquad S_{\rm nc}(x) = \frac{L}{\ell} \sum_{\mu > \nu} F^2_{\mu\nu}(x),
\eeq{ncqed}
does not favor the presence of monopoles in the continuum
limit since the flux on each plaquette is peaked around zero when
one takes $L\to\infty$ at a fixed $\ell$.  The compact Wilson
gauge action,
\beq
{\cal W}_{\rm g}\equiv e^{-\sum_x S_{\rm c}(x)};\qquad S_{\rm c}(x) =\frac{2L}{\ell} \sum_{\mu > \nu} \left [ 1-\cos\left( F_{\mu\nu}(x)\right) \right],
\eeq{cqed}
on the other hand, does not suppress monopoles in the continuum
limit since the flux $F_{\mu\nu}(x)$ has multiples peaks around $2
\pi N_{\mu\nu}(x)$ with integer values of $N_{\mu\nu}$ -- monopoles
are counted per cube~\cite{DeGrand:1980eq} by writing $F_{\mu\nu}(x)
= \bar F_{\mu\nu}(x) + 2\pi N_{\mu\nu}(x)$ where $\bar F_{\mu\nu}(x)
\in [-\pi,\pi)$ and $N_{\mu\nu}(x)$ are integers. The monopole
charge inside a cube with one corner at $x$ is given by
\be
Q(x) = \frac{1}{2}\epsilon_{\mu\nu\rho} \nabla_\mu N_{\nu\rho}(x)\label{mondef}.
\ee
The Villain gauge action~\cite{Villain:1974ir,Drouffe:1983fv},
\beq
{\cal W}_g\equiv \sum_{\{N_{\mu\nu}\}} e^{-\sum_x S_{\rm v}\left(x,\{N_{\mu\nu}\}\right)};\qquad  S_{\rm v}(x) = \frac{L}{\ell} \sum_{\mu> \nu} \left( F_{\mu\nu}(x) -2\pi N_{\mu\nu}(x)\right)^2
\eeq{vqed}
is also a compact action but has the advantage that the integer
part of the flux per plaquette is made explicit.  We have introduced
new degrees of freedom $N_{\mu\nu}(x)$ and one needs to sum over
all integer values to define the partition function.  This action
is expected to be in the same universality class as the compact
Wilson gauge action.  The Villain action allows for all values of
$Q(x)$ with the only condition that the sum over all $x$ in a finite
lattice with periodic boundary conditions will be zero.  The only
coupling in all cases is $\ell$ which can be viewed as the dimensionless
extent of the lattice and the lattice spacing is $a=\frac{\ell}{L}$.

One can only consider the part of the above Villain action restricted
to the sector $Q(x)=0$ for all $x$.  If the manifold is ${\mathbf
R}^3$, then this automatically implies that
\beq
N_{\mu\nu}(x)=\nabla_\nu
n_\mu(x)-\nabla_\mu n_\nu(x),
\eeq{zeroq}
for integers $n_\mu$. On $T^3$, as used in Monte Carlo simulations,
the condition in \eqn{zeroq} implies $Q(x)=0$ but further restricts
the sum of $N_{\mu\nu}$ on any $(\mu\nu)$-plane to be zero. In
particular, this disallows configurations with net constant flux
$2\pi {\cal Q}$, for integer ${\cal Q}$, over any of the $(\mu\nu)$-plane
in the continuum limit.  However, such an extra restriction on $T^3$
cannot be important in the thermodynamic limit since any equal and
opposite fluctuations in flux in different parts of the lattice are
allowed. For values of $N_{\mu\nu}$ of the form in \eqn{zeroq}, one
can change $\theta_\mu(x)\to\theta_\mu(x)-2\pi n_\mu(x)$ and annul
the term $N_{\mu\nu}$. Therefore, the Villain path integral restricted
to values of $N_{\mu\nu}$ of the type in \eqn{zeroq} is the same
as the standard non-compact QED$_3$ path integral defined using
\eqn{ncqed}.  Similarly, one can constrain the integer valued flux
$N_{\mu\nu}$ to take a particular value $N^{Q\overline{Q}}_{\mu\nu}$
defined via
\beq
\frac{1}{2}\epsilon_{\mu\nu\rho}\nabla_\mu N^{Q\overline{Q}}_{\nu\rho}(x)=Q\delta_{x,y}-Q\delta_{x,y'}.
\eeq{mmbar}
The above constraint corresponds to an insertion of flux $Q$ monopole
at a lattice site $y$ and a flux $Q$ antimonopole at $y'$, and this
cannot be absorbed by a change of variable of the gauge fields.
The monopole correlator in nc-QED$_3$ can simply be defined as the
ratio of path integrals subject to the constraint in \eqn{mmbar}
with $Q=1$ to that with $Q=0$~\cite{Peskin:1977kp}.  Instead, we
find the gauge field background $ {\cal A}^{Q\overline{Q}}_\mu(x)$ that minimizes
\beq
S_{\rm v}^{Q\overline{Q}} = \sum_{x,\mu < \nu} \left( B^{Q\overline{Q}}_{\mu\nu}(x) -2\pi N^{Q\overline{Q}}_{\mu\nu}(x)\right)^2;\qquad
B^{Q\overline{Q}}_{\mu\nu}(x) = \nabla_\nu {\cal A}^{Q\overline{Q}}_\mu(x)-\nabla_\mu {\cal A}^{Q\overline{Q}}_\nu(x),
\eeq{min}
and couple the theory to this classical background field
in order to define
\beq
Z_Q=\left(\prod_{x,\mu}\int_{-\infty}^\infty d\theta_\mu(x)\right) {\det}^{N/2}\left[\slashed{C}^\dagger(U) \slashed{C}(U)\right] 
e^{-\frac{L}{\ell}\sum_{y,\mu>\nu}\big{[}F_{\mu\nu}(y)-B^{Q\overline{Q}}_{\mu\nu}(y)\big{]}^2}.
\eeq{zq}
The advantage of using $B^{Q\overline{Q}}_{\mu\nu}$ over using $2\pi
N^{Q\overline{Q}}_{\mu\nu}$  is that background field coupling has
no effect in pure gauge theory, and any effect that is observed in
$Z_Q$ will arise only due to the presence of fermions. 
This
follows from a simple change of variable $\theta_\mu(x)\to \theta_\mu(x)-{\cal A}^{Q\overline{Q}}_\mu(x)$ that eliminates ${\cal A}^{Q\overline{Q}}_\mu(x)$
only in the case of pure gauge path integral. As we already noted, 
$N^{Q\overline{Q}}_{\mu\nu}(x)$ cannot be written as a curl, and hence
such a change of variable is not possible even in pure gauge theory.
On ${\mathbf
R}^3$, the resulting ${\cal A}^{Q\overline{Q}}$ is the field for a
Dirac monopole-antimonopole pair.  The advantage of minimizing
\eqn{min} on toroidal lattice is to take care of both the lattice
discretization as well as the periodicity correctly.  We checked
through a full fledged computation in the case of $N=2$ QED$_3$
that the difference in $Z_Q/Z_0$ between the minimum on the torus
as defined above and the discretized field of Dirac monopole-antimonopole
pair as defined in~\cite{Karthik:2018rcg} is, however, marginal.

Lets denote the lattice distance between the monopole and antimonopole
as $T=|y-y'|$, which is related to the physical separation $\tau=T
a$. Then, the ``bare" monopole-antimonopole correlation function
in lattice units, $G^{(Q)}_{\rm B}$, is the ratio of partition
functions with and without the flux $Q$ monopole-antimonopole
insertion~\cite{Peskin:1977kp}:
\beq
G^{(Q)}_{\rm B}(\tau,\ell,a)=\frac{Z_Q}{Z_0};\qquad \tau=T\frac{\ell}{L}.
\eeq{correl}
Our specific choice for the location of the monopole and anti-monopole
in \eqn{mmbar} is realized by
\beq
N^{Q\overline{Q}}_{12}(0,0,x_3)=2\pi Q;\qquad 1\le x_3\le T,
\eeq{nfield}
and zero for all other directions and lattice points $(x_1,x_2,x_3)$.
The square tube with non-zero integer flux running between the
monopole at $y=(0,0,0)$ and $y'=(0,0,T)$ is the Dirac string.  Any
other configuration for this Dirac string that is simply connected
to the above construction is related through appropriately chosen transformations
$\theta(x)\to\theta(x)+2\pi n(x)$.  The details pertaining to the
construction of the background field $ {\cal A}^{Q\overline{Q}}_\mu$
can be found in~\cite{Karthik:2019jds}.

\section{Method and simulation details}
\bef
\centering
\includegraphics[scale=0.8]{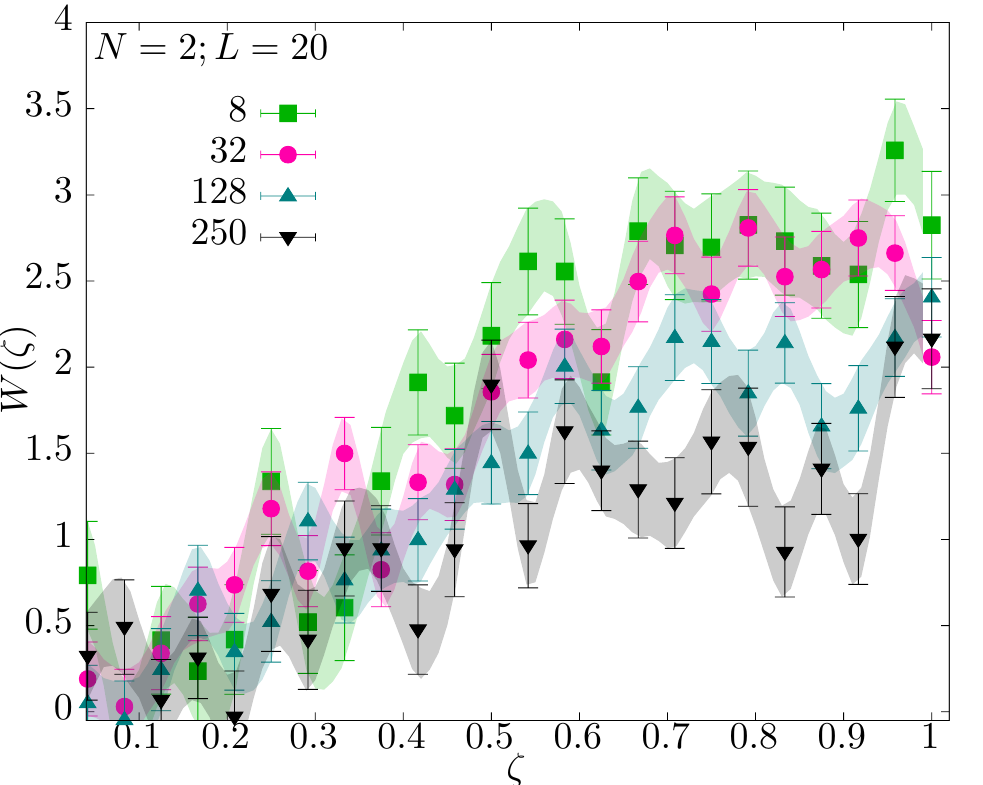}
\caption{
    $W(\zeta)$,  is shown as a function of $\zeta$ at different 
    values of $\ell$ at fixed $L=20$ for the case of $N=2$ flavors. 
    The different colored symbols correspond to different physical
    extents $\ell$, and the bands are the cubic spline interpolation
    of the data points.
    The free energy for $Q=1$ monopole-antimonopole pair 
    is given by the area under the curves, $\int_0^1 W(\zeta) d\zeta$.
}
\eef{wzetanf2}
In a Monte-Carlo simulation, it is only possible to compute ensemble
averages and not the partition function itself. A brute force way to
implement the correlator in \eqn{correl} is to compute
the average
\beq
G^{(Q)}_{\rm B}(\tau,\ell, a)=\left\langle e^{\frac{L}{\ell}\sum_{y,\mu>\nu}B^{Q\overline{Q}}_{\mu\nu}(y)\left(2F_{\mu\nu}(y)-B^{Q\overline{Q}}_{\mu\nu}(y)\right)}\right\rangle_{0},
\eeq{brute}
where $\langle\cdots\rangle_0$ is the ensemble average with respect
to $Z_0$ for $N$ flavor theory. The problem with such an approach
is the absence of overlap between the configurations sampled by
$Z_0$ and $Z_Q$.  In order to avoid this overlap problem, we couple
QED$_3$ to the background gauge field $\zeta {\cal A}^{1\overline{1}}$
through a generalization of \eqn{zq} to non-integer values of $Q$,
where $\zeta$ is a tunable auxiliary variable~\cite{Kajantie:1998zn}.
Consistent with the previously introduced notation, the resulting
partition function is $Z_\zeta$. From this, we can compute the
lattice free-energy ${\cal F}^{(Q)}_{\rm B}(\tau,\ell,a)$ to introduce
the monopole-antimonopole pair separated by physical distance $\tau$
in an $\ell^3$ torus at finite lattice spacing $a$ as,
\beq
{\cal F}^{(Q)}_{\rm B}(\tau,\ell,a)\equiv -\log\left[G^{(Q)}_{\rm B}(\tau,\ell,a))\right]=\int_0^Q d\zeta W(\zeta),
\eeq{freecomp}
where,
\beq
W(\zeta)=\frac{-1}{Z_{\zeta}}\frac{\partial Z_{\zeta}}{\partial\zeta}=\frac{2L}{\ell}\bigg{\langle} \sum_{x}(F_{\mu\nu}(x)-\zeta B^{(1)}_{\mu\nu}(x))B^{(1)}_{\mu\nu}(x)\bigg{\rangle}_\zeta.
\eeq{wdef}
Thus, $W(\zeta)$ can be computed in the Monte-Carlo simulation of
$Z_{\zeta}$ through the measurement of $F_{\mu\nu}(x)-\zeta
B^{(1)}_{\mu\nu}(x)$ on the gauge fields that are sampled. In this
paper, we will only study $Q=1$ monopoles and we drop labels for
$Q$ henceforth.

A way to determine the correlator in \eqn{correl} is to compute
$G_{\rm B}(\tau,\ell,a)$ at different large values of $\tau$ in an
$\ell^3$ box. At each fixed $\tau$, one should first convert the
lattice correlator to a renormalized physical one, then take the
continuum limit $L\to\infty$ at a fixed $\ell$, followed by the
infinite volume limit $\ell\to\infty$. Finally, one can consider
the asymptotic $\tau\to\infty$ limit to study its $\tau^{-2\Delta}$
scaling.  However, such a method is not practical since it requires
computations of multiple values of $\tau$ per Monte Carlo sample
point in the parameter space, and further introduces unwanted
systematic errors from the $\ell\to\infty$ extrapolations at fixed
$\tau$.  As was demonstrated in the case of monopole
correlators~\cite{Karthik:2018rcg}, a better method is to make use
of scaling of correlators near the infrared fixed point.  That is,
one expects the scaling
\beq
G_{\rm B}(\tau,\ell,a)= a^{2d} G_{\rm R}(\tau,\ell); \qquad G_{\rm R}(\tau,\ell) = \frac{1}{\ell^{2\Delta}}{\cal G}\left(\frac{\tau}{\ell}\right),\quad \text{as}\ \tau,\ell\to\infty.
\eeq{scale1}
The conversion factor $a^{2d}$ takes the bare correlator to the
renormalized correlator of the naive dimension $d$ monopole
operator~\footnote{The symbol $d$ should not be confused with the
Euclidean space-time dimension which is always 3 in this paper.}.
The subtle issues with this will be addressed in the next section.
In addition, the leftmost expression is only true up to finite $a$,
or equivalently finite $1/L$, corrections.  Assuming, we have
obtained the renormalized correlator, the second expression exhibits
its scaling near the infrared fixed point.  We do not have to make
any further assumption about the form of ${\cal G}(\tau/\ell)$ if
we fix $\tau/\ell=\rho$ as $\ell$ is varied. Here, we take $\rho=1/4$.
Equivalently, the free energy to introduce a monopole-antimonopole
pair separated by distance $\tau=\rho\ell$ would be
\beq
{\cal F}_{\rm R}(\ell)=-\log\left[G_{\rm R}(\rho\ell,\ell)\right]=f_0(\rho)+2\Delta \log(\ell),
\eeq{logscl}
up to higher-order corrections in $1/\ell$. Since we keep $\rho$
fixed in this paper, we keep its dependence implicit. It will be
useful to consider the free-energy per two-component flavor as
\beq
f_{\rm R}(\ell)\equiv \frac{{\cal F}_{\rm R}(\ell)}{N}=f'_0(\rho)+\frac{2\Delta}{N}\log(\ell).
\eeq{fpern}
In the limit of $N\to\infty$, both $f'_0(\rho)$  and $\Delta/N$
have well-defined limits. In $1/N$ expansion, one finds
\beq
\frac{\Delta}{N}=\Delta^{\infty}+\frac{k}{N}+\ldots,
\eeq{onebyn}
with $k<0$. The large-$N$ value $\Delta^{\infty}$ was computed using
free fermion coupled to monopole background since it was argued
that the fluctuations in dynamical gauge fields are suppressed by
$1/\sqrt{N}$. Such an analysis gave
$\Delta^{\infty}=0.265$~\cite{Borokhov:2002ib,Pufu:2013eda,Karthik:2018rcg}.
For the $Q=1$ monopole we consider here, the leading correction was
computed to be $k=-0.0383$~\cite{Pufu:2013vpa}.

\bef
\centering
\includegraphics[scale=0.62]{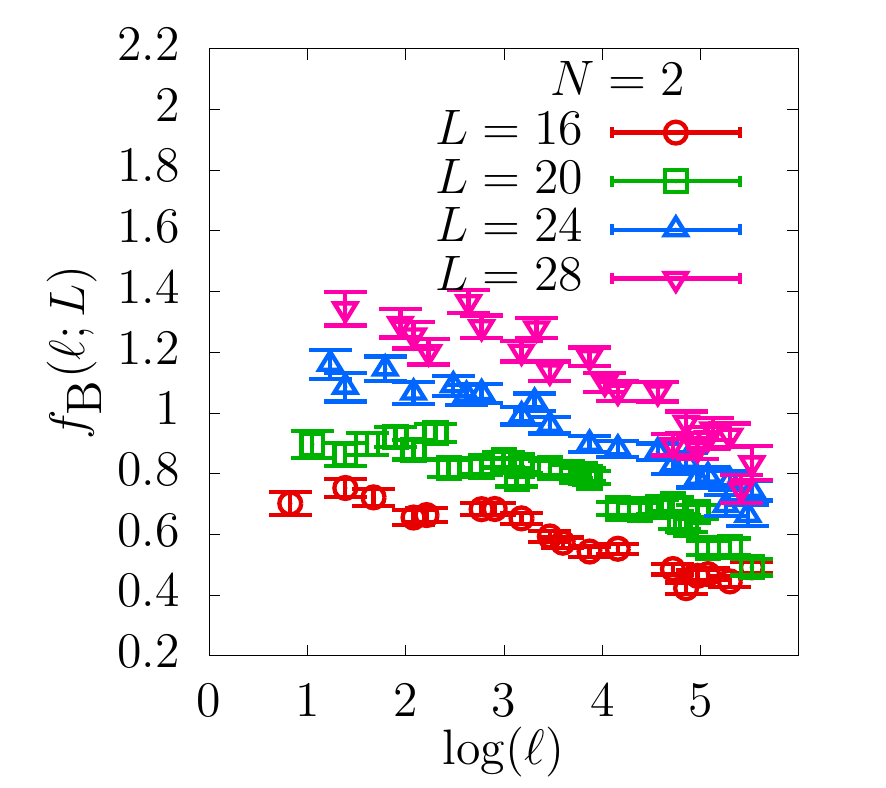}
\includegraphics[scale=0.62]{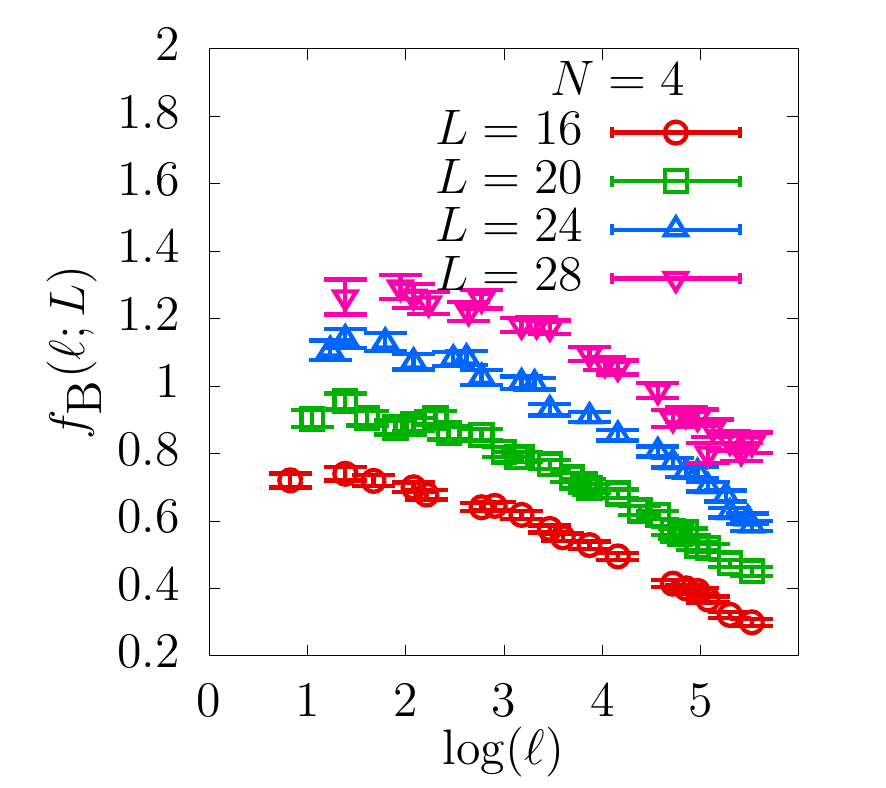}
\includegraphics[scale=0.62]{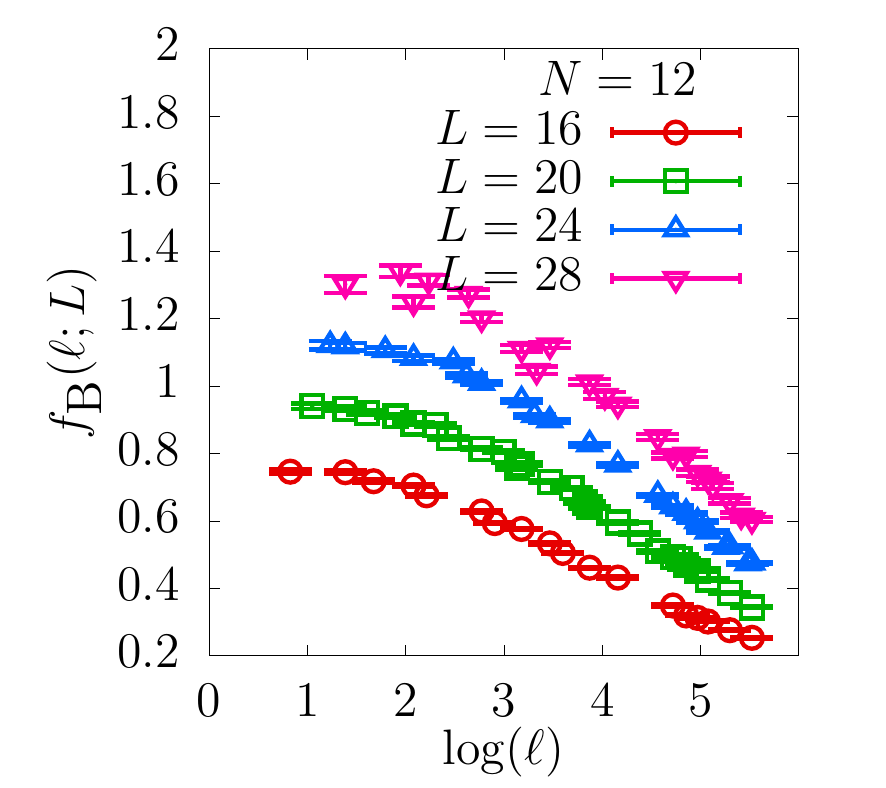}
\caption{
    The bare free energy per fermion degree of freedom, $f_{\rm B}={\cal F}_{\rm B}/N$, of  the lattice monopole-antimonopole background
    field insertion is shown as a function of physical extent of
    the box $\ell$. 
    The three panels from left to right correspond to $N=2,4$ and 12
    respectively. The different colored symbols correspond to 
    different $L$ specified in the key.
}
\eef{barefree}

In the current work, we studied $N=2,4$ and 12 flavors of fermions
--- the idea being that we can use $N=12$ is to check for consistency
with large-$N$ expectations, and use $N=2,4$ to study the effect
of smaller $N$.  We sampled configurations from $Z_\zeta$ using 50K
trajectories of hybrid Monte-Carlo (HMC) simulation. For each value
of $\ell$, $L$ and $N$, we simulated 24 different equally spaced
values of $\zeta$ from 0 to 1. At each $\zeta$, we computed $W(\zeta)$
using Jack-knife analysis to take care of autocorrelation, and
performed the numerical integration in \eqn{freecomp} after smoothly
interpolating the 24 data points for $W(\zeta)$. We used different
values of $\ell$ ranging from $\ell=1$ to $\ell=250$ at each fixed
values of $L$. To estimate the continuum limit of the $\ell$
dependence of the free energy, we used $L^3$ lattices with $L=16,20,24$
and 28. In \fgn{wzetanf2}, we show $W(\zeta)$ as determined for
$N=2$ at four different values of $\ell$ on $20^3$ lattice as a
sample.  The area under each of those curves gave the bare free energy
${\cal F}_{\rm B} = -\log G_{\rm B}$.

\section{Results}

\bef
\centering
\includegraphics[scale=0.62]{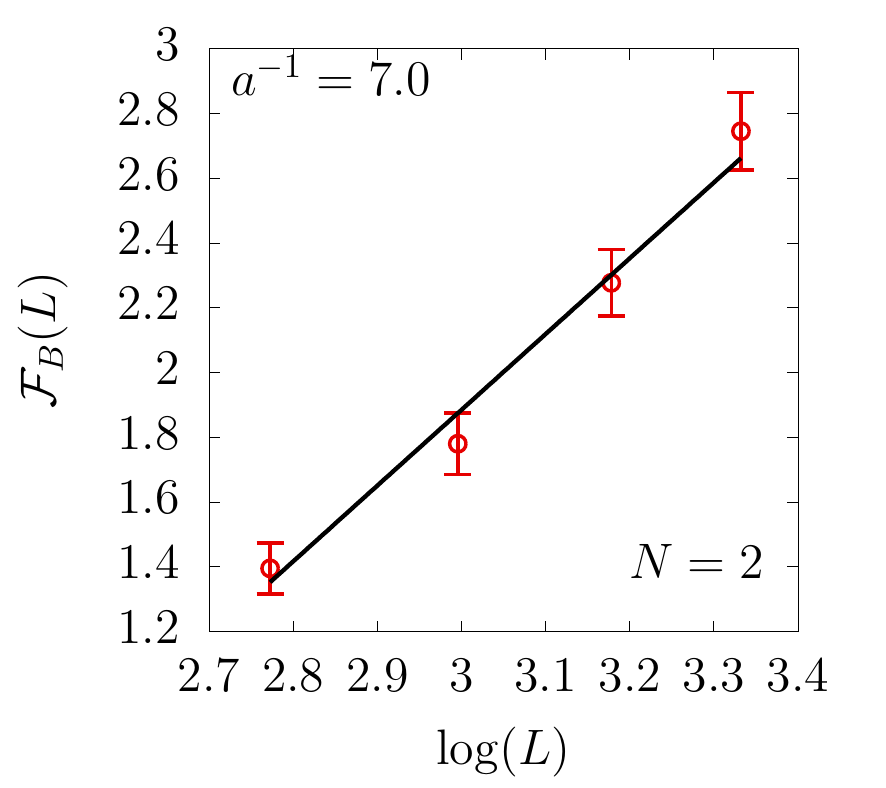}
\includegraphics[scale=0.62]{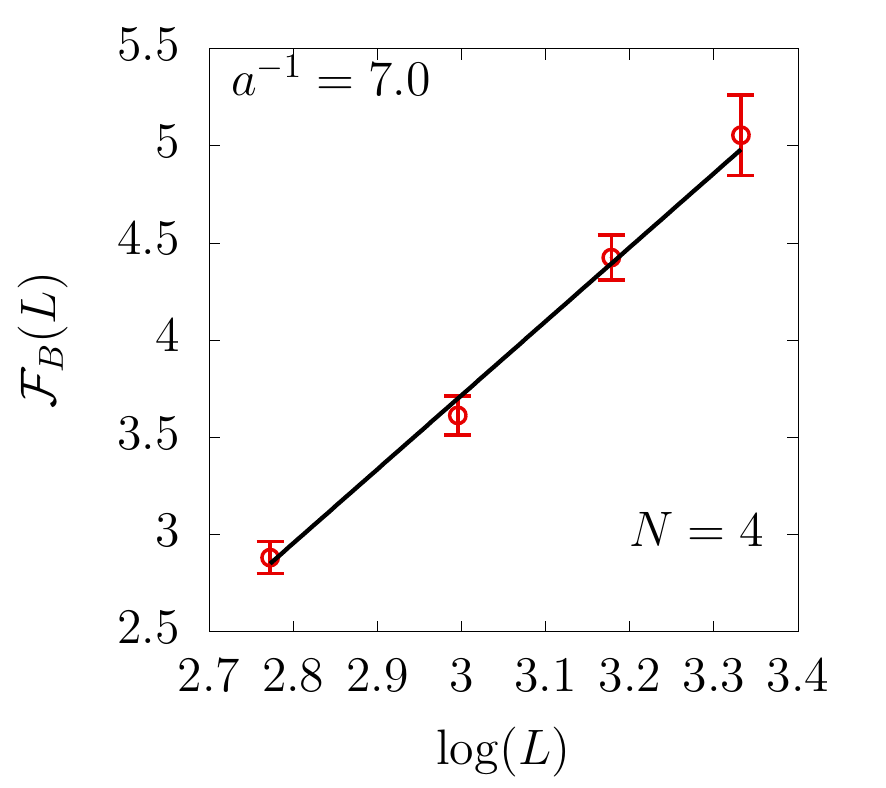}
\includegraphics[scale=0.62]{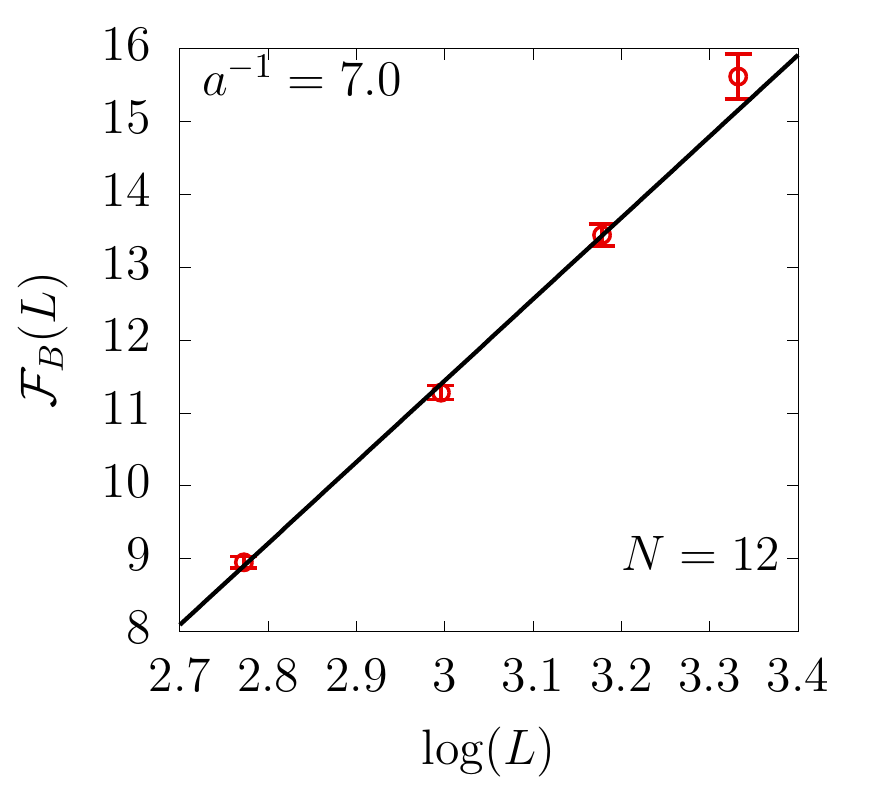}
\caption{
    Determination of monopole naive dimension $d(L)$ as determined
    in the range of $L=16, 20, 24$ and 28 by a linear fit of ${\cal
    F}_{\rm B}$ at fixed lattice spacing corresponding to $a=\frac{1}{7}$
    to an effective $\log(L)$ dependence over the range of $L$
    considered.  The three panels from left to right correspond to
    $N=2,4$ and 12 respectively.
}
\eef{scaledim}

First, we show the dependence of the lattice free energy per flavor,
$f_{\rm B}(\ell,L)=N^{-1}{\cal F}_B(\rho\ell,\ell,\ell/L)$, for
introducing monopole-antimonopole pair at distance $\tau=\ell/4$
from each other on the box size $\ell$ in \fgn{barefree} for different
fixed values of $L$.  The plots from left to right are for $N=2,4$
and 12 flavors of two-component fermions respectively. This dependence
as computed using $L=16,20,24$ and 28 are shown as different colored
symbols.  As expected, the bare lattice free energy from different
$L$ do not fall on a universal curve since the lattice spacing
$a=\ell/L$ keeps changing as $\ell$ is varied at fixed $L$.  In
fact, as it stands the result seems unphysical --- the free energy
decreases with increasing $\ell$ at fixed $L$.  Therefore, we have
to first convert the lattice correlator $G_{\rm B}$ to the correlator
in physical units, $G_{\rm R}$ by determining $d$ in~\eqn{scale1}.
Since QED$_3$ is super-renormalizable, $d$ for a local operator
would simply be its naive dimension (e.g., the flavor triplet vector
operator ${\cal O}_V$ with $d=2$).  However, defining the monopole
correlator through background field coupling is different at least
in two ways --- (A)~Even at the Gaussian fixed point (i.e., $\ell=0$
at finite $L$), the monopole correlator defined in \eqn{correl}
scales as $L^{-2d}$ only asymptotically as $L\to\infty$.  One needs
to contrast this with the correlator of ${\cal O}_V$ at the Gaussian
FP, which would scale as $L^{-4}$ for all $L$.  In other words, the
background field method singles out the scaling operator of lowest
naive dimension only in the large-$L$ limit at the Gaussian FP.
(B)~The effective action for the background field is apparently
non-local, and numerically showing that it can be renormalized by
a simple $a^{2d}$ factor is non-trivial.

As discussed above, for the background field coupling at finite $L$
at $\ell=0$ (or $a=0$), one can only obtain an $L$-dependent effective
value of the scaling dimension, $d(L)$, which we expect to approach
the free fermion value $d$ in the limit of very large $L$ that are
not feasible in the computation presented here.  The value of $d(L)$
relevant in the range of $L$ studied here can be determined numerically
via $L^{-2d(L)}$ fit to $G_{\rm B}(\rho L a,L a,a)$ in the limit
of $a\to 0$ over a small range of $L$.  With this $L$-dependent
value of $d$ determined at the Gaussian fixed point, we can best
approximate the renormalized correlator at other non-zero $\ell$
and $a$ by using
\beq
G_{\rm R}(\rho\ell,\ell)=\left(\frac{\ell}{L}\right)^{-2d(L)}G_{\rm B}(\rho\ell,\ell,a),
\eeq{dim}
which automatically ensures that the correlator $G_{\rm R}(\rho\ell,\ell)$
has no $L$ dependence for $\ell\approx 0$. For larger $\ell$, the
residual $L$ dependence in $G_{\rm R}(\rho\ell,\ell)$ is a lattice
artifact, which can be removed by $L\to \infty$ continuum extrapolations.
To actually study QED$_3$ in the strict $\ell\to 0$ limit at finite
$L$, one needs to integrate the fermion determinant over the still
unsuppressed constant modes of the gauge fields in all three
directions of the torus. This is nontrivial to implement, and hence
we consider the result from the fixed, small $a=\frac{1}{7}$ as an
approximation of the strict $\ell=0$ results.  In \fgn{scaledim},
we have shown the $\log(L)$ dependence of the free energy ${\cal
F}_{\rm B}(\rho L a, L a,a)$, simply denoted as ${\cal F}_{\rm
B}(L)$, at this fixed $a=\frac{1}{7}$ over the range of $L$ used
in this paper. This corresponds to changing $\ell=112$ at $L=16$
to $\ell=196$ at $L=28$ in each of the panels in \fgn{barefree}.
It was possible to fit the data over the range of $L$ from 16 to
28 using $f_0+2d\log(L)$ and thereby obtain the value of $d(L)$
over this range of $L$.  The fits are shown as the black straight
lines in the three panels. We obtained the slope $2d(L)$ as 2.34(22),
3.81(21) and 11.17(48) for $N=2$,4 and 12 respectively,   which
correspond to $d(L)/N$ of 0.585(56), 0.476(53) and 0.465(20)
respectively.  Since the value of $G_{\rm B}$ at $\ell=0$ depends
on the distribution of constant gauge fields in the three directions
that are allowed at $\ell=0$ for any finite $N$, the value of
$d(L)/N$ for intermediate $L$ can depend on $N$.  Instead of the
above method, where $\ell$ is varied at fixed $L$ thereby forcing
us to construct $G_{\rm R}$ from $G_{\rm B}$, we could have instead
studied $G_{\rm B}$ at fixed lattice spacing. However, achieving
larger physical volumes at fixed small lattice spacing would become
numerically prohibitive.

\bef
\centering
\includegraphics[scale=0.85]{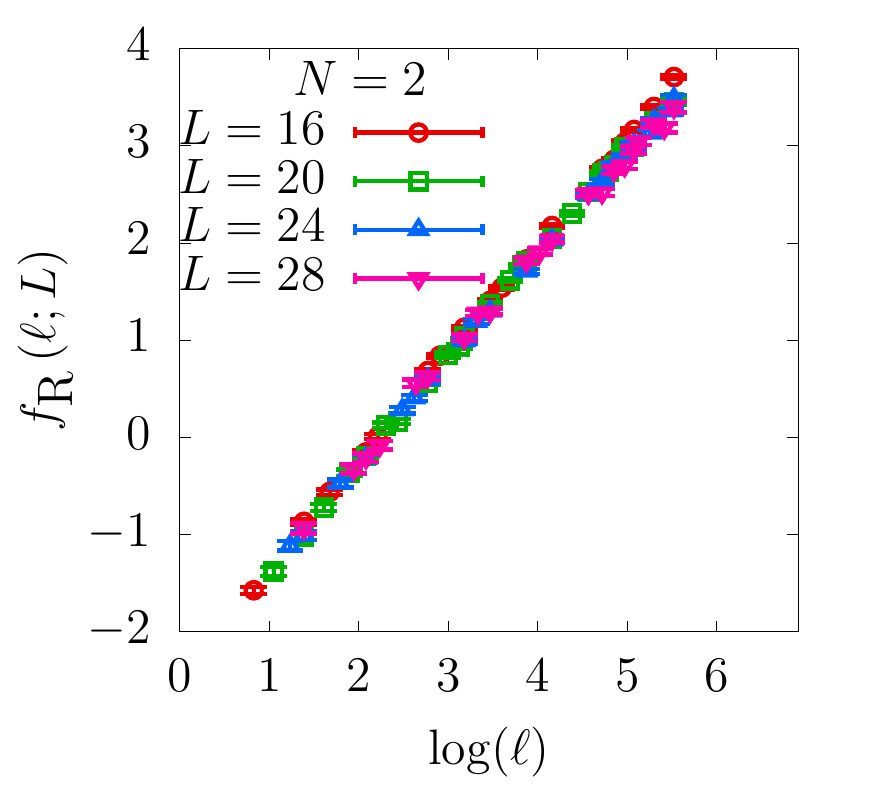}
\includegraphics[scale=0.85]{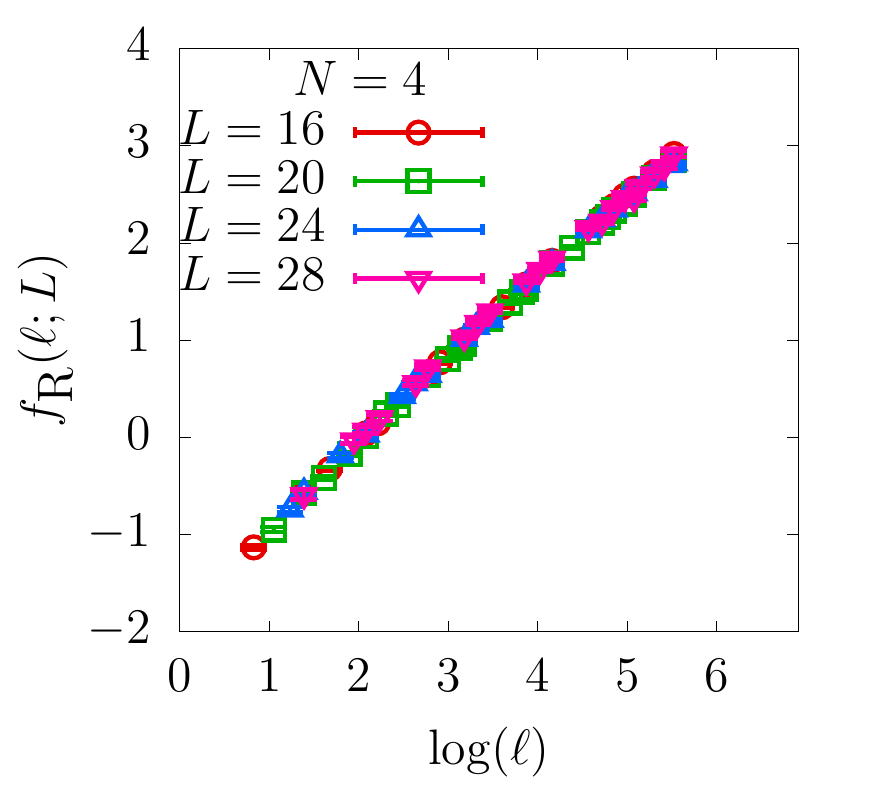}
\includegraphics[scale=0.85]{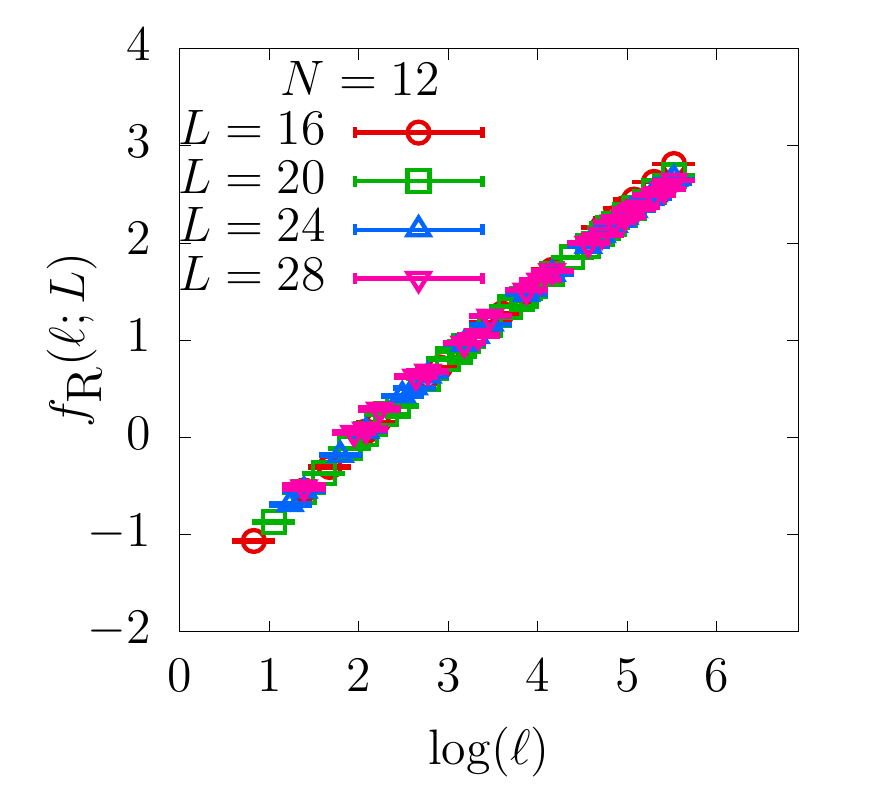}
\caption{
    The physical free energy per two-component fermion flavor,
    $f_{\rm R}(\ell)$, required to introduce a monopole-antimonopole
    pair of Dirac string length $\ell/4$ is shown as a function of
    the box size $\ell$. The top-left, top-right and the bottom
    panels are for $N=2,4$ and 12 respectively. The data as obtained
    from different $L$ are distinguished by the different colored
    symbols.
}
\eef{renormfree}
In \fgn{renormfree}, we show the resulting free-energy per two-component
flavor, $f_{\rm R}(\ell)=N^{-1} {\cal F}_R(\rho\ell,\ell)$. This
was obtained by adding $2d(L)\log\left(\frac{\ell}{L}\right)$ to
${\cal F}_{\rm B}(\rho\ell,\ell,\ell/L)$  and then computing the
resulting renormalized free energy per two component fermion.  The
three panels are for the three different values of $N$. The data
from different $L$, made distinct by the colored symbols, now fall
on near universal curves. This data collapse is quite non-trivial
and supports the assumption that we have defined the correlator of
an operator that has a local description in the continuum. Contrary
to the behavior of the bare lattice free energy, the renormalized
free energy starts increasing with $\ell$ as physically expected
since one does not expect monopoles to be spontaneously created in
non-compact QED$_3$. For all $N$, including $N=2$, the dependence
of $f_{\rm R}(\ell)$ shows no evidence of a linear $\ell$ dependent
piece corresponding to an exponential fall, $G_{\rm R}(\ell)\sim
\exp(-\mu\rho\ell)$, with a mass $\mu$ that could set the scale for
a scale-broken theory.  Assuming QED$_3$ with $N=2,4,12$ flow to
infrared fixed points as $\ell\to\infty$, the asymptotic values of
slope in this linear-log plot would give the values of $2\Delta(N)$.
As it can be seen, the slope changes with $\ell$ and extracting the
value of $\Delta$ will require extrapolations.  Therefore, first
we focus on model-independent inferences from the data.

\bef
\centering
\includegraphics[scale=0.9]{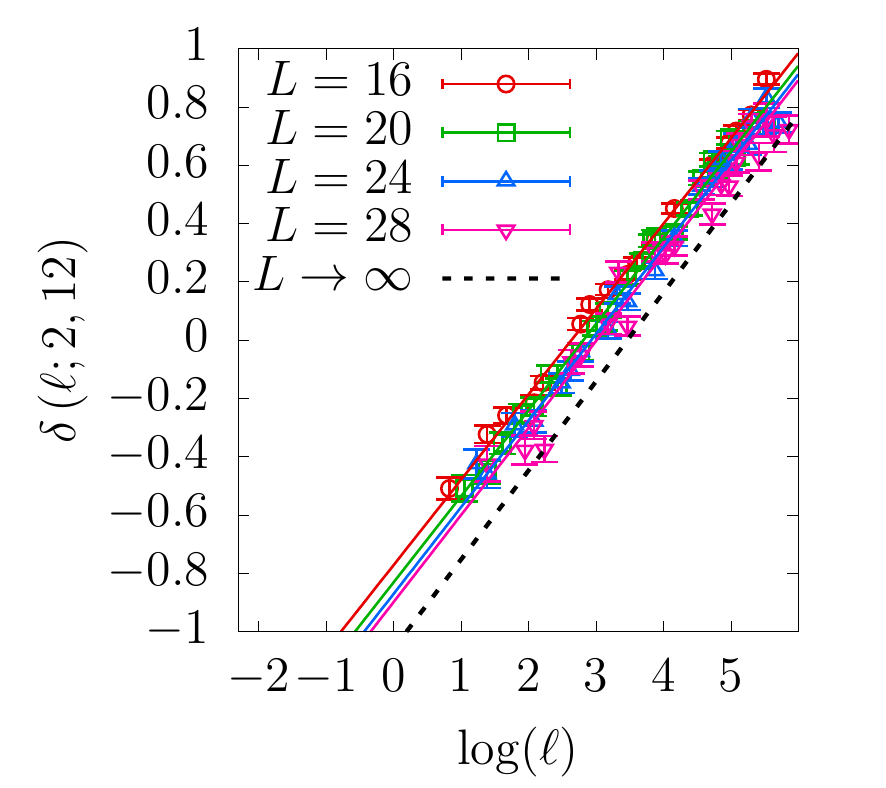}
\includegraphics[scale=0.9]{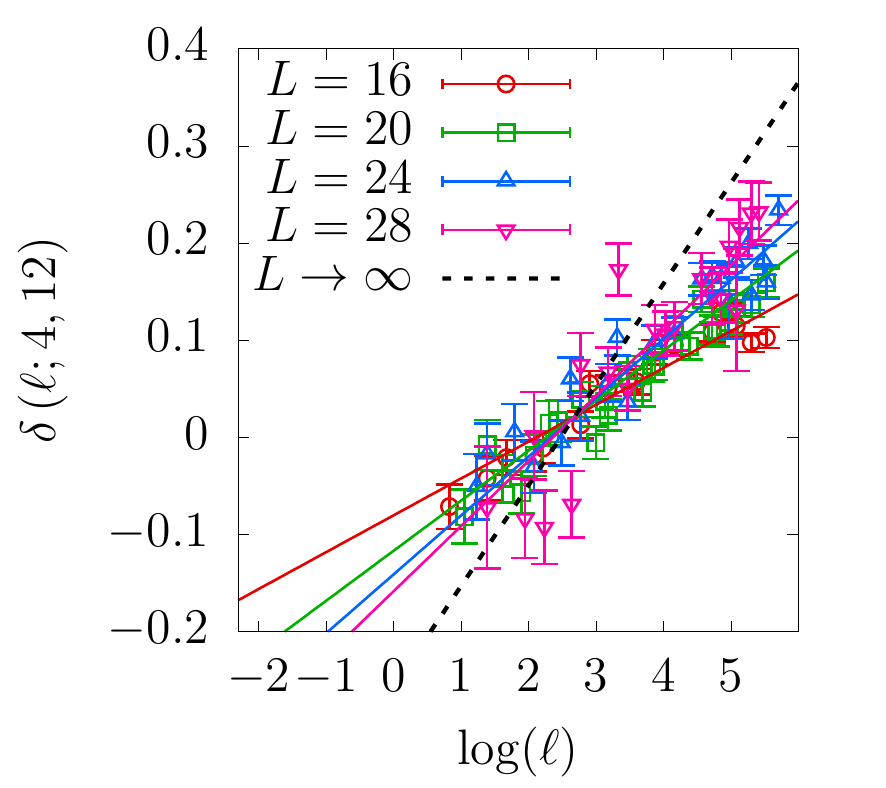}
\caption{
    The difference $\delta\left(\ell;N,N'=12\right)=f_{\rm
    R}(\ell;N)-f_{\rm R}(\ell;N'=12)$ is shown as a function of
    $\ell$ for $N=2$ and 4 in the left and right panels respectively.
    The solid lines are the combined fits to the form
    $\delta\left(\ell;N,12\right)=a_0+a_1/L+2\left(b_0+b_1/L\right)\log(\ell)$.
    The dashed line is the central value of the estimated continuum
    limit, $L\to\infty$, of this difference.  The positive slope,
    both in the data as well as in the estimated continuum limit,
    clearly indicates that $N=2$ and $N=4$ have larger infra-red
    scaling dimension than that of $N=12$.
}
\eef{dfree}

From the three panels in \fgn{renormfree}, where we have kept the
range of $\ell$ and $f_{\rm R}$ in the plots to be the same, we find the free
energy per flavor $f_{\rm R}(\ell)$ for $N=12$ shows a weaker
dependence on $\ell$ compared to $N=2,4$.  To make this quantitative,
in \fgn{dfree}, we show the difference
\beq
\delta(\ell;N,N')=f_{\rm R}(\ell;N)-f_{\rm R}(\ell;N'),
\eeq{diffdef}
between the free energy per flavor in $N$ and $N'$ flavor
theories~\footnote{The dependence on $N$ which is implicit in
$f_R(\ell)$ is explicitly shown in the notation used in \eqn{diffdef}.}.
In the infra-red, we expect such a difference to be
\beq
\delta(\ell;N,N')=2\left[\frac{\Delta(N)}{N}-\frac{\Delta(N')}{N'}\right]\log(\ell),
\eeq{diffree}
as $\ell\to\infty$.  For $N'$, we choose the largest value $N'=12$,
that we have.  We have shown $\delta(\ell;2,12)$ in the left panel,
and the difference $\delta(\ell;4,12)$ in the right panel as functions
of $\log(\ell)$.  At leading order in $1/N$, this difference vanishes.
Instead, we find that both $\delta(\ell;2,12)$ and $\delta(\ell;4,12)$
increases with $\ell$ making it quite evident that $\Delta(N=2)/2$
and $\Delta(N=4)/4$ are larger than $\Delta(N=12)/12$.  This effect
is arising purely due to finite value of $N$.  Quite surprisingly,
$\delta(\ell;N,12)$ shows a logarithmic dependence on $\ell$ over
the entire range of $\ell$ used. Perhaps, this is due to the finite
$\ell$ corrections to the infrared scaling get approximately canceled
between $f_{\rm R}$ at $N$ and $N'$. Therefore, we performed a
combined fit to the data for $\delta(\ell;N,12)$ at all $\ell$ from
different $L$ using an ansatz,
\beq
\delta(\ell;N,12)=\left(a_0+\frac{a_1}{L}\right)+2\left(b_0+\frac{b_1}{L}\right)\log(\ell),
\eeq{ans1}
with $a_0,a_1,b_0$ and $b_1$ as fit parameters. The value of $b_0$ will then
give us an estimate of the difference $\Delta(N)/N-\Delta(12)/12$ in the
continuum limit $L\to\infty$. We find such a fit ansatz to describe the 
data for both $N=2$ and 4 well with $\chi^2/{\rm dof}<2$ with 81 data
points in each fit. 
We find the best fit parameters $[a_0,a_1,b_0,b_1]$ 
for $N=2$ and $N=4$ to be $[-1.067(77),4.7(1.4),0.1531(94),-0.11(18)]$ and 
$[-0.262(52),2.92(98),0.0529(61),-0.54(11)]$ respectively. The
different $L$-dependence of the $N=2$ and 4 data is due to an empirical
$N$-dependence of the
coefficients $a_1$ and $b_1$ for the finite $1/L$ corrections.
The resulting
best fits as evaluated at $L=16$,20,24 and 28 are shown by the straight lines 
in the linear-log plots with the same colors as that of the corresponding
data. The black dashed lines in the two panels are
the estimates of the continuum limit of $\delta(\ell;N,12)$. We find from the 
value of $b_0$ that 
\beqa
\frac{\Delta(2)}{2}-\frac{\Delta(12)}{12}&=&0.153(9);\cr
\frac{\Delta(4)}{4}-\frac{\Delta(12)}{12}&=&0.053(6).
\eeqa{est1}
In the expansion in $1/N$ around the large-$N$ fixed point, the
leading contribution to the difference is
$\frac{\Delta(N)}{N}-\frac{\Delta(N')}{N'}=k\left(\frac{1}{N}-\frac{1}{N'}\right)$
with $k=-0.0383$. Thus, to order $1/N$, the value of scaling dimension
decreases at finite $N$ from the large-$N$ value, and the slope of
$\delta(\ell;N,N')$ as a function of $\log(\ell)$ should be negative
at this order. Our numerical result for $N=2$ and 4 on the other hand,
suggests the opposite behavior for the values of $N$ which are
${\cal O}(1)$.  This implies that higher order terms in $1/N$ that
are of opposite sign cannot be ignored, or perhaps a breakdown of
the $1/N$-expansion.  However, it is true that the corrections
themselves are also small.

\begin{figure}
\centering
\includegraphics[scale=0.9]{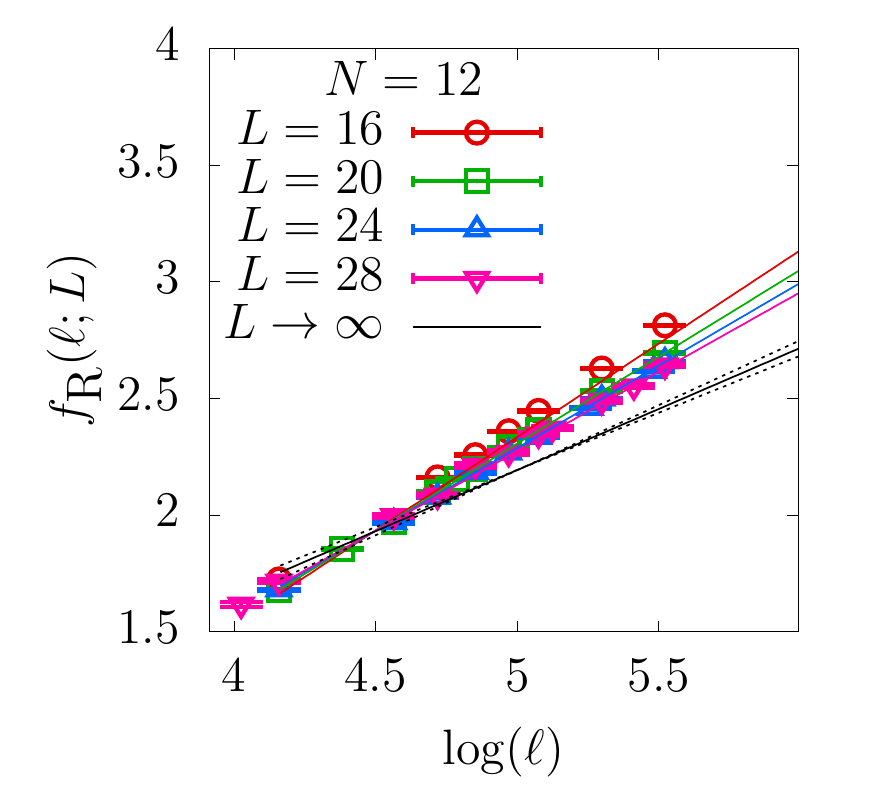}
\caption{
    The large $\ell$ behavior of $f_{\rm R}$ for $N=12$ is shown.
    The different colored symbols are for different $L$.  The solid
    straight lines are from combined fits of the form $f_{\rm
    R}(\ell)=\alpha_0+\alpha_1/L+2(\beta_0+\beta_1/L)\log(\ell)$
    to the data at $L=16,20,24,28$ from $\ell=64$ to 250. The back
    straight line and band are the estimates for the  $L\to \infty$
    continuum limit. The estimate for $\beta_0$ is $0.26(2)$, and
    can be identified with $\Delta(12)/12$.
}
\label{fg:largen}
\end{figure}

Now, it remains to be shown that for $N=12$, for which one might
naively expect the large-$N$ expansion to hold, and hence the
asymptotic value of the scaling dimension $\Delta(12)$ is consistent
with large-$N$ expectation. Unlike the above conclusion about the
correction to large-$N$ behavior, extraction of $\Delta(12)$ requires
modeling and extrapolations since the free-energy does not exhibit
a pure $\log(\ell)$ dependence over the entire range of $\ell$ used
in this computation.  In \fgn{largen}, we focus only on values of
$\ell> 64$. By fitting a simple ansatz
\beq
f_{\rm R}(\ell)=\left(\alpha_0+\frac{\alpha_1}{L}\right)+2\left(\beta_0+\frac{\beta_1}{L}\right)\log(\ell),
\eeq{ans2}
to the $N=12$ data for $f_{\rm R}(\ell)$ over the larger values of
$\ell$ from different $L$, we estimate the value of $\Delta(12)/12$
as the best fit value of $\beta_0$. The best fit $\log(\ell)$
dependence for $L=16$, 20,24 and 28 are shown along with the data
in \fgn{largen}, 
that are described by $[\alpha_0,\alpha_1,\beta_0,\beta_1]=[-0.42(18),-19(4),0.261(19),2.22(41)]$. 
Though the data seems to be well described by such
an ansatz, the $\chi^2/{\rm dof}$ is about 3 due to the much smaller
errors in the data for $N=12$.  The black dashed line is the estimated
$\log(\ell)$ dependence in the $L\to\infty$ limit. We estimate the
slope of this continuum dependence as $\beta_0=0.26(2)$. For
comparison, the value of $\Delta(12)/12$ from large-$N$ up to leading
order in $1/N$ is 0.262. Our estimated value of $\Delta(12)$ is
consistent with this value, and thereby lends further support for
our numerical work.  This implies that the monopole scaling dimension
for $N=12$ is estimated to be 3.24(24). This is consistent with
$N=12$ being the critical flavor where the $Q=1$ monopole operator
becomes just marginally relevant in the infrared fixed point.

\section{Conclusion and discussion}

In this paper, we presented an ab initio lattice computation of the
monopole correlator in $N=2,4$ and 12 flavor massless QED$_3$ by
using the background field method.  To avoid the overlap problem
which would make the computation of ratio of partition functions
with and without monopole-antimonopole background field, we slowly
increased the value of monopole flux from from 0 to integer $Q$.
One of noteworthy result in this paper is the feasibility of this
approach itself, seen via the good signal to noise ratio of the
monopole free-energy (which is the negative logarithm of the monopole
correlator). This encourages the application of this method to other
QFTs where monopole operators can be defined.  We demonstrated
empirically that the monopole correlator behaves like a local
operator and  can be simply ``renormalized" by the factor $a^{2d}$
at lattice spacing $a$, where $d$ is the naive monopole dimension
as obtained on $L^3$ lattice in the limit $a\to 0$. The key numerical
result for the free-energy to introduce monopole-antimonopole pair
in $N=2,4$ and 12 flavor QED$_3$ is shown in \fgn{renormfree}. Using
this data, we demonstrated that the scaling dimension for $N=12$
QED$_3$ is consistent with large-$N$ expectation and that $N=12$
is consistent with being the critical flavor where monopole scaling
dimension takes the marginally relevant value of 3. By computing
the differences in monopole free-energy between $N=2$ and 12, and
between $N=4$ and 12, we found evidence for the deviation of the
scaling dimension in $N=2,4$ theories from the $N=12$ theory to be
positive. This is in contradiction to the $1/N$ analysis up to
${\cal O}(1/N)$. It remains to be seen if this tension can be
resolved by inclusion of higher-order corrections in $1/N$ in the
analytical expressions, or points to a breakdown of the $1/N$
framework itself where in the fixed point for smaller $N$ belongs
to a different family than the one in large-$N$. In conclusion, the
results in the paper along with slightly different analytical results
from large-$N$ analysis supports the direct computation of $N$
flavor compact QED$_3$ around $N\approx 12$, which however requires
algorithmic development to deal with large number of near-zero
Wilson-Dirac modes.

In the paper, we did not demonstrate in the lattice regularization
framework that monopoles carry flavor quantum number and breaks the
U$(N)$ flavor symmetry to U$(N/2)\times$U$(N/2)$ symmetry.
Demonstrating this is not important to the computation presented
in this paper, but central to the U$(N)$ flavor symmetry breaking
in $N_c$ flavor compact QED$_3$.  In the continuum, one
shows~\cite{Dyer:2013fja,Borokhov:2002ib} this by noting that the
ground-state of the Hamiltonian of massless fermion on $S^2$ with
constant flux $2\pi Q$ has $QN$ zero modes. Thus, the gauge-invariant,
CP-invariant vacua are obtained by filling the $QN$ zero modes with
$QN/2$ fermions picked amongst $N/2$ flavors and $QN/2$ antiparticles
picked amongst another $N/2$ flavors.  Thus, the vacua with monopole
background transform under the irreps of the U$(N)$ flavor symmetry.
With the lattice regularized fermions on spherical monopole background,
we do not have a similar derivation to study the flavor structure
of the vacuum. A difficulty is defining the lattice fermion on
$S^2$. Therefore, to gain an understanding of similar flavor symmetry
breaking mechanism on the lattice, we consider a background field
on $T^3$ corresponding to constant flux $2\pi Q$ on all $T^2$ spatial
slices, as an analogue of constant flux background on $S^2\times
R$ for the spherical monopole. It was shown~\cite{Karthik:2015sza}
that the two-dimensional transfer matrix for the two-component
Wilson-Dirac operator, $\slashed{C}_W$, has $2L^3+Q$ eigenvalues
greater than 1 and $2L^3-Q$ eigenvalues less than 1 for $m_w>0$; a
consequence of gauge field topology in two dimensions.  Thus the
vacuum has total $Q$ fermions. Similarly, for $\slashed{C}^\dagger_W$
with $m_w<0$, the vacuum has total $Q$ anti-fermions.  With $N/2$
fermions coupled via $\slashed{C}_W$ and another $N/2$ fermions
coupled via $\slashed{C}^\dagger_W$, the vacuum has $N/2$ fermions
from $N/2$ flavors and $N/2$ anti-fermions from the other $N/2$
flavors. This is very similar to the structure seen in a spherical
monopole background. The flavor symmetry-breaking on this particular
background is determined by the need to preserve parity: a choice
of $N/2$ flavors coupling to $\slashed{C}_W$ and the other ones to
$\slashed{C}^\dagger_W$.  We expect similar mechanism to be true
for massless Wilson-fermion on monopole background.

\acknowledgements

R.N. acknowledges partial support by the NSF under grant number
PHY-1515446. N.K. acknowledges support by the U.S. Department of
Energy under contract No. DE-SC0012704. N.K. thanks the physics
department at IMSc, Chennai for their kind hospitality during the
course of preparing this manuscript. This work used the Extreme
Science and Engineering Discovery Environment (XSEDE), which is
supported by NSF under grant number ACI-1548562.  Resources at the
San Diego Supercomputer Center were used under the XSEDE allocation
TG-PHY180011.

\bibliography{../../mynotes/biblio}

\end{document}